# CAN GRAPHENE BILAYERS BE THE MEMBRANE MIMETIC MATERIALS? ION CHANNELS IN GRAPHENE-BASED NANOSTRUCTURES

**Oleg V. Gradov**

Institute of Energy Problems of Chemical Physics, Russian Academy of Sciences, http://www.inepcp.ru
117829 Moscow, Russian Federation
gradov@chph.ras.ru

**Margaret A. Gradova**

Semenov Institute of Chemical Physics, Russian Academy of Sciences, http://www.chph.ras.ru
119991 Moscow, Russian Federation
Institute of Energy Problems of Chemical Physics, Russian Academy of Sciences, http://www.inepcp.ru
117829 Moscow, Russian Federation
gradova@chph.ras.ru

*Abstract.* **The prospects of application of graphene and related structures as the membrane mimetic materials, capable of reproducing several biomembrane functions up to the certain limit, are analyzed in the series of our papers. This paper considers the possibility of the ion channel function modeling using graphene and its derivatives. The physical mechanisms providing selective permeability for different membrane mimetic materials, as well as the limits of the adequate simulation of the transport, catalytic, sensing and electrogenic properties of the cell membrane ion channels using bilayered graphene-based structures are discussed.**

CONTENTS

## 1. INTRODUCTION

In [1] provides an overview of membrane models – semi-synthetic, synthetic, biogenic, semiconducting, superconducting, ferroelectric - and the related membrane mimetic materials from phospholipid bilayers and Langmuir films to multilayer nanofilms and ferromagnetic structures, along with the consideration of the problems of ambiguity models, completeness the membrane mimetic materials modeling, functional and substrate equivalence of the membrane mimetic materials. The use of analogical functional similarity criteria for the analysis of possibility of consideration of graphene, particularly the two-layer graphene as the membrane mimetic material - the subject of this work, with a focus on the analysis of it key property – permeability for ions, water and other organic matter.

## 2. SEMIPERMEABILITY AND THE ION CHANNELS

A well known flexibly adjustable semipermeability towards various agents is characteristic for the graphene-oxide membranes in organic solvents, with the size of the nanochannels providing either transport or blocking of several agents, which can be narrowed down by the thermal annealing or extended by solvation, thereby changing the boundary of the transmitted agent's size selectivity [2]. This corresponds to the well known models of the "size / solubility" – regulation of the sodium ion channels [3] and good approximations of the kinetic regimes depending on the ionic radius [4]. As a particular case of the applicability of the quantitative relations between the structure and biological activity (QSAR) [5], the analysis principles for the functional adjustability of the ion channels by the sink parameters, with the membrane pores corresponding to the drains, and the size of particles penetrating through those pores, in the case of the graphene layers fail to describe the membrane permeability, since the membrane considered performs a biological function despite the fact that its chemical composition is far from bioorganic one, and hence, is not included into the ion channel databases [6] which could be used to obtain QSAR data.

It is noteworthy that similar problems with the QSAR approaches earlier emerged in the analysis of the lipid nanopores operating as the ion channels of the cell membranes without any conventional ion channel components [7] due to their size corresponding to the ion radius. (Typically even at the dependence of the several ion channel group input on the ionic radius complex biochemical and crystallographic models are applied up to the homohexamer one [8], i.e. compatible with the QSAR principles based on the analysis of the conformation and steric accessibility of the channelome biomacromolecules).

In connection with the above contradiction it is also worth to mention the known data on the ionic permeability of the synthetic filters which have never been analyzed using QSAR methods for the similar reasons: it was shown that the ions penetrating through the synthetic PET filters are selected by the mechanisms providing the biomimetic and membrane mimetic kinetics with the discrete fast transitions between the conductivity levels and parametric selectivity of the ion fluxes, as well as inhibition by divalent cations, similar to the cell ion channels [9, 10].

Thus, it is possible to work out the criteria for verification of the biomimetic nature of the membrane mimetic models, based on the idea about the gradations of the ion channel efficiency according to the conductivity and bioelectric response generated, rather than on the conventional statement about the discrete transition between the excited states («all-or-none law» [11, 12]), typical for the deterministic model of the electrophysiological potential generation.

It is reasonable, since it has been shown earlier that the ion channel states (which are the prototypes of the membrane mimetic, or even channel mimetic models [13-15]) with a low ion conductivity in fact are not "closed", but are changing within several gradations providing qualitatively ("ion-selectively") and quantitatively (according to the permeability) distinguishable types of the membrane-electrophysical or electropysiological response, which is the main cause for the specific channelome noises at the patch-clamp registrograms [16]. Of course, this does not mean the close similarity between the biological and synthetic membrane structures or the possibility to reduce the channelome molecular machinery to the simple porous structure, but it indicates the general operation principles of the non-specific physical mechanisms at the nanoscale where the size effects are physically-determined and chemically-independent, i.e. are little if any influenced by the chemical composition of the medium and macromolecular morphology of the biopolymers (in a conventional meaning of this term introduced by P.J. Flory [17, 18], a Nobel prize winner and the author of the Flory-Huggins equation which is formally similar to the Van der Waals equation, and hence, can be applied at the physical scale considered in this paper).

For the above reasons, biomimetic interpretation and assigning of the specific membrane properties to graphene and other layered membrane mimetic materials with the pore size corresponding to the ion radii, although does not contradict to the experimental data, in fact is a logical error of the inversion type, arising from the mixing of the deductive and inductive predicates: graphene and the cell membrane with the similar pore size obviously subject to the same physical and chemical principles (which is manifested in the similarity of the solvation regulation and the electrophysical response), and hence, in this aspect "graphene can be considered as a membrane mimetic material to the same extent as the biological membrane can be considered as a graphene-mimetic".

This suggests that the similarity problem (including the above mentioned QSAR) for such structures, if considered from the standpoint of molecular biology rather than biophysical and biocolloid chemistry, becomes incorrect. According to the classical colloid and capillary chemistry principles [19, 20] it is possible to determine the efficiency limits of the sorption processes within the pores and channels of the particular diameter, which is usually applied in cytophysiology [21] (the Freundlich adsorption isotherm [22] is named after Herbert Freundlich – the founder of the capillary chemistry and a pioneer of the capillary-chemical approaches in physiology). Thus, it is more appropriate to address the problem of channelomics of the graphene-oxide membrane mimetic materials from the standpoint of the size effects inherent to the former in the aspects of adsorption, filtration and size separation of the particles /ions, rather than in the framework of the specialized (supra)molecular structures which satisfy the similarity conditions to the specific cytophysiological structure performing the physical and chemical separation mechanism (i.e. "a similarity of the general phenomenon to its particular manifestation").

The effects of the size-dependent penetration of the chemical agents through the graphene layers are well known in nanochemistry [23, 24]. Reactivity towards different gases [25] and electrochemical properties of the nanoelectrode graphene layers [25] also depend on the pore diameter. The above phenomenon known as "size-dependent electrochemistry" is a conceptual continuation of the Freundlich's "Kapillarchemie". This approach is widely applicable not only in the chemistry of the carbon structures, but also in the nanostructural materials science as a whole.

Regarding the carbon structures related to graphene, nanoelectrochemistry based on the size effects was applied in the studies on the multilayered carbon nanotubes [27] and a fully or partially reduced graphite oxide [28] which is an electrode material for electrochemical double-layer capacitors [29] often considered as the biological membrane bilayer model [30-34], as well as the layered supercapacitors of EDLC-type ("electric double-layer capacitor") with the pronounced relation between the ion and pore sizes [35] similar to the biological membranes and membrane mimetic materials considered above. By the way, this is a reduced graphite oxide which is used as an electrode material in such supercapacitors [36], allowing the EDLC-based membrane models to mimic not only the energy storage function, but also a number of biochemical receptor functions, such as recognition and detection of the specific mediators/neurotransmitters/neurohormones (e.g. dopamine – [37-39]), performed on the basis of the electrochemical principles without any specific biomacromolecular agents [40-42] (in particular, due to the thickness-dependent hydrophobic properties of graphene, which is formally similar to the size-dependent properties of the "trans-graphene transport" of differently charged biomolecular agents through the pores in its hydrophobic surface [43, 44], which makes the problem of designing the graphene-based sensitive biomimetic materials soluble and similar to the design of the receptor mimetic peptides [45] based on the hydrophobic interaction simulations).

To date, all the exceptions from this reductionistic rule, actually, confirm the rule, since the range of charge, mass and other physical and chemical parameters of the substances used in the composite techniques as well as the degrees of their biochemical (or immunochemical in the case of such detection principles) affinity differ by orders of magnitude: aptamers/nucleic acids [46, 47]; conductive polymers, such as polypyrrole (both at graphene and pyrolytic graphite) [48-51]; porphyrins and their derivaives, qualitatively different in their physical and chemical properties and aggregation behavior in solutions [52-54]; aminosugars - linear polysaccharide derivatives, such as chitosan [55, 56] used for immobilization; polymer electrolytic membranes, particularly those based on fluoride containing copolymers – fluorocarbon vinyl esters containing sulfonic groups (e.g. a well known nafion), including those with the composite impregnation by several inorganic components and structure-modifying agents [57-60]; inorganic particles and clusters themselves – gold, copper, nickel and zinc oxide nanoparticles [61-68]; graphene-doping chemical elements, such as nitrogen [69, 70]. Although the above list is not complete, it fully represents the range of different mass and charge parameters for a number of molecules in the membrane mimetic structures performing receptor functions based on graphene and its derivatives. Despite the fact that in some cases electrophysical response has not been directly registered (detected only from the secondary indicators, such as redox-dependent fluorescence), the above information can be sufficient to prove by contradiction the correctness of the reductionistic model which does not require any bioorganic or macromolecular agents for performing the receptor functions by the membrane mimetic materials.

It is noteworthy that graphene-based nanostructures are often used in designing of both n-layered supercapacitors and sensors to various agents operating without enzymes

[71]. In such sensors graphene is only a kind of supercapacitor material – there are biosensors [72], gaze [73] and humidity [74] sensors (also capable of the energy storage) based on the non-graphene supercapacitor membrane coatings. Wherein, from the standpoint of the channelomics of such membrane mimetic materials, it should be mentioned that supercapacitor sensor membranes can separate ions, synthetic and biological molecules [75], promote dissociation of salts producing energy [76] due to the ion gradient (similar to the above cited Mitchell chemiosmotic model). Thus, energy supplying and sensing properties of the graphene-based membrane mimetic materials, which are phenomenologically similar to a number of fundamental properties inherent to the biomembranes, are provided by the electrophysical mechanisms (the molecule polarity, hydrophobicity and a number of properties related to the uncompensated charge interactions, in particular, coordination). It is not reasonable to consider the biosensor properties of the graphene-based membrane mimetic materials in this section, since they will be considered in details below, as well as the chemosensor ones, but the sensor properties of the supercapacitor-based membrane mimetic materials were discussed here for the reason of extrapolation of the ion channel- / membrane mimetic modeling criteria and the search for the general basis of the sensor, ion-selective and electrophysical properties of the model system.

The aforementioned correlations between the electric and transport (ion-selective) properties require the example of their cooperative interaction. To date there are well known graphene-based ion-selective field-effect transistors (ISFET) sensitive to the proton concentration [77], which clearly correlates to the chemiosmotic model of the membrane electrogenesis, as well as with the model of biomembrane as the transistor structure known since 1960-th [78]. In such models the ion transport, sensing properties and changing of the sensor electric parameters are synchronized and interdependent. By the way, for all chemically-selective field-effect transistors (ChemFET), and particularly for the ion-selective field-effect transistors (ISFET) [79, 80], there is no difference between the "membrane selectivity" and the "sensor electric response specificity" [81]. In recent years the ion sensing functions in microfluidic and even nanofluidic chemometric systems (labs-on-a-chip) are performed by the ionic transistors based on the ion concentration polarization by an ion exchange membrane [82], capable of performing electrogenic ion exchange and sensing with the electric response. Similar functions can be performed by the graphene or graphene oxide-based ion exchangers and composite ion-exchange materials [83-85], whereby the principles of synchronization of various aspects of their activity as multifunctional membrane mimetic materials can be realized using graphene and its analogs/derivatives applicable as FET. It should be noted that for performing of the most of the above mentioned functions graphene-based structures should posses only the FET properties, but not the whole ChemFET. There are sensors based on the graphene-containing FET structures (non-positioned as a ChemFET), but applied for the redox-sensing [86], aptasensing [87], electrochemical biosensing [88], metal ion sensing [89], as well as for design of the components of the bioelectronic "nose" [90]. This is due to the qualitative dependence of the graphene-based FET response on the ambient liquid and vapor media conditions [91], which may be due to the increased mobility of the charge carriers (holes), the decrease in the residual charge carrier concentration and changes in the molecular transport, induced by the charged defects (so-called perforation) near the surfaces of the active layers. The above effects result from the charge transport, as well as the molecule orientation induced by the charge near the FET surface [92], but it requires the presence of the vacant pores and charge defects in the membrane mimetic surface similar to the structural and electrophysiological features of biomembranes with the electrogenic properties, ionic transport and spatial molecule orientation in the controlling electric field are coupled and colocalized [93]).

## 3. PORES AND ION CHANNELS IN GRAPHENE-BASED NANOSTRUCTURES

Let us consider here the question about the presence of the pores and the ion channel analogs (or the possibility of their realization) in graphene and its products, precursors and related compounds. There are known gaze-transport [94] and ion-transport [95] channels in the laminar graphene oxide and graphene-based nanostructures of the electrophysical destination used in electrotechnics for designing of the lithium-ion batteries. All the phenomena of the selective ion penetration through the graphene structures are based on the above channel operation. In literature one can find the description of all kinds of the ion channels implemented on the basis of the graphene nanopores/nanochannels [96, 97]: for alkali metals [98], including biomimetic (bioinspired) analogs of the sodium and potassium channels [99] (as well as the other alkali metals, such as lithium, rubidium, and caesium [100-102], as follows from the thermodynamic calculations for the monovalent cations [103] and quantum chemical simulations for a number of the alkali metal ions [104], which can also operate in the channelome, but their clark is too low, and hence, does not allow to perform any significant functions at the macroscopic scale at the organism or the whole biosphere level), for the chloride ions [105] (the analogs of the known chloride ion channels [106, 107]), etc.

Consideration of the "exotic" pores and channels in graphene operating with the rubidium, gadolinium and other rare earth element ions [108, 109] is beyond the scope of this paper for the purpose of maintenance of the biomimetic functional analogy.

In the synthesis of the biomimetic graphene-based ISFET structures an important aspect is the compatibility with the channelome, since the carbon nanostructure-based ISFET are used in the electrophysical studies in neurocytology [110] by contact formation with the neurons, which serve as the ion-exchange structures between the channelome and FET, and are responsive to the extracellular medium and the mediated stimulation/inhibition of the ion-transport channel activity [111]. Recent data on the possibility of the biomimetic ion channel function modeling and their introduction into the graphene capsules and membrane-like structures, by 2015 made it possible to consider the graphene-based capsules with the ion-selective channels as the embryo protocells [112, 113]. Though the above interpretation is doubtful, given that the early protocell models were actually simple phospholipid membrane structures [114-116], a membrane mimetic nature of this abstraction is clear, while it is not a breakthrough among the numerous non-lipid [117], membrane-free [118] and electrostatically-gated [119], inorganic [120, 121] mineral [122, 123] and other protocells, illustrating the recent tendency to the biomembrane substitution by its functional alternative.

However, it is evident that the synthetic functional model operates differently from its prototype, while providing chemical sensing in the vicinity of the graphene surfaces [124]: selective particle transport through the graphene layers can be performed accordingly not only to the particle charge or mass [125], but also to its spin [126], which is fundamentally different from the conventional physics of the biomembrane prototypes (although spin labels and other methods of the spin chemistry are often used in biophysics and cytophysiology for studying the ion channel properties and membrane permeability [127-131]). Thus, here we speak about the transduction of the physical agents rather than chemical (ionic) carriers of the QSPR/QSAR-coupled chemical parameters and physiological properties. In this case a model can be considered objective only if it is adequate to the biological prototype by the formation mechanism; functionally different models are based on different principles, and hence, reproduce the properties of the prototype only to the extent determined by the difference in their formation mechanisms.

There are known works on the graphene nanofluidic channel formation by scrolling graphene layers into a tube [132]. In biological systems there is an analogous example of the model channel formation via self-assembly (folding) of the cyclic peptide nanotubes [133] formed directly in the course of their interaction with the ligand [134] (as a supramolecular response [135] to the above interaction). Similarly occurs self-assembly of the model ion channel networks based on liquid-crystalline bicontinuous cubic phases [136] or columnar phases based on crown ethers in lipid bilayers [137]. The ion channel assembly from dendrimers is slightly different due to their branched structure [138], but this special case can not be implemented using graphene membrane mimetic structures, and hence, is beyond the scope of our paper. Another different mechanism of the membrane self-assembly, and hence, the membrane channel assembly, occurs under the templating conditions, which is simulated by the formation of the graphene films on various catalytic and ultramicrostructured surfaces (e.g. for obtaining FET and other ion-selective structures [139, 140]).

Catalytic structuring of the subsurface layer [141] is an inevitable condition for the on-surface synthesis. Templating on the inorganic catalytic surfaces is essential for the synthesis of a variety of inorganic catalytically-active (to a certain extent of self-assembly even autocatalytically-active) redox-surfaces [142]; similar requirements are in the synthesis of the layers with channels, mediated by the templating metal surfaces [143]. In such syntheses vacancies serve as the precursors (seeds) for the channel formation and are also involved in the determination and direction of the surface structuring forms under the phase transitions [144]. A similar but qualitatively opposite role is played by the metal ions: surfactant templating on the molten salt surfaces with the metal adducts leads to the formation of the metallotropic liquid-crystalline phases [145]. Given the applicability of the catalytic templating methods for the synthesis of the graphene-oxide nanostructures [146], it is possible to design graphene-based biomimetic /membrane mimetic surfaces with the ion channel function performed by the structures formed via templating and related mechanisms. This will correspond both to the template-associated synthesis of the peptide ion channel-mimicking systems [147] and to the formation of the synthetic inorganic transmembrane tubes and channels under the lipid templating [148], i.e. it will be substrate-independent from the biomolecular and supramolecular carriers, as well as from the organic/inorganic composition of the ion channel-mimicking structure as a whole, which is a prerequisite for the agent modeling of their functional mimetics.

Addressing the problem of catalysis in the ion channel mimetic self-assembly, including graphene-based structures, it is necessary to point out the catalytic function of the channels-prototypes. Catalytically-active are both cationic (e.g. calcium channels [149], characterized by the coupling of their catalytic and transport functions in the ATP-mediated $Ca^{2+}$-transport), and anionic [150] ion channels and membrane ion pumps. Many specific regulators of the ion transport and permeability, such as CFTR – cystic fibrosis transmembrane conductance regulator [151, 152], catalytic and the channel opening/closing regulation functions are also coupled. ATP-sensitive potassium channels (sometimes referred to as KATP/KATP or sarcKATP in sarcolemma, mitoKATP in chondriome, nucKATP at the nucleus depending on their localization), also possess catalytic activity [153, 154]. A similar situation occurs with the redox-regulators and the iron metabolism mediators – ferritin-based ion channels [155], as well as with their synthetic derivatives and analogs made from nanostructures [156].

In general, synthetic catalytically-active pores [157] with the ion selectivity similar to their biological prototype are capable of performing the ion channel functions with the same (bio) catalytic function. The possibility of the above principle implementation using the agent models is determined by several conditions. Firstly, both for the potassium [158] and sodium [159] channels with the opposite operation modes (potassium channels are opened while the sodium ones are closed, since the cell resting potential parameters are close to the Nernst equilibrium potential of the potassium ions) catalysis plays an important role. Secondly, catalytic functions are characteristic both for the cationic and anionic channels (the term "chloride ion channels" in this case is not fully appropriate, since the same channels provide transport of the $HCO_3^-$, $I^-$, $SCN^-$ and $NO_3^-$ anions). Thirdly, most of the model ion-selective systems are equally sensitive to most of the monovalent cations [160]. On the other hand, catalytic functions and the ligand recognition can be performed almost without any ion channels based on the pi-cation interactions [161]. At the same time coupling between the potential generation phases and catalytic cycles in the membrane can be provided by the lipids, e.g. by a phospholipid PI(4,5)P2 (phosphatidylinositol 4,5-bisphosphate) [162, 163]. Thus, the nature and structure of the agent itself are not significant for mimicking its function, while the adequate function reproduction is the key similarity criterion of the biomimetic model to its biological prototype.

Another characteristic example is the operation of the water channel associated with the catalytically-active sites [164] and their mimicking within the graphene layer or graphite surface in the form of the water transpiration channels [165, 166], which can be reproduced (with the presence of the driving gradient) beyond the structural modeling and design/synthesis of the ion channel mimetics by means of the simple passive membrane model with its permeability dependent on the ion size, and water permeability within the same channels and the surface features [167]. This corresponds to the membrane pore model and the model of the non-selective independent ion channels with their permeability determined by the ion and molecule size. However, the native water channels – integral proteins aquaporins – also perform the pore functions, while some of them depending on the molecular size and shape also allow penetration of glycerin, ammonia, urea and carbon dioxide through the membrane [168].

## 4. CONCLUSION

In a general case, agent functional modeling of the ion channels using graphene-based structures is not only possible, but also satisfies the requirement for the colocalization of the ion-selective and electrogenic functions, resulting from the analysis of the biophysical prototype (membrane) functions. Good examples of such structures with colocalization are the channels in the graphene-based ISFET – ion-selective field-effect transistors [169, 170] considered above. However, there are two aspects of electrogenesis, which can not be neglected during the analysis of the ion channel operation in graphene-based ISFET. One of them is the electrical double layer, which is the absolute theoretical limit of the nanoelectric system design, and the other one is the double layer nature of the biological membrane as a capacitor (which is required for modeling its periodical discharge in the form of the action and breakdown potentials in the area of the membrane pore formation), which should correspond to the double-layered structure of the graphene-based agent membrane mimetic material. Then the logic "performance" and the "duty cycle" of the graphene ion channel operation [171] in the electrogenic medium

– graphene bilayer – will possess a higher degree of similarity with its biological prototype. A detailed consideration of the electrical double layer as a driving force of the electrogenic processes and membrane mimetic structures, along with the consideration of the differences between the double-layered and multi-layered graphene-based structures and their single-layer analogs in the membrane mimetic aspect will be given in the next part of this paper.